\documentclass[10pt,superscriptaddress,secnumarabic,nobibnotes,aps,prl,showpacs,twocolumn,nofootinbib]{revtex4-2}%

\usepackage[OT1]{fontenc}
\usepackage[utf8]{inputenc}
\usepackage{bm}
\usepackage{mathrsfs}
\usepackage{subfigure}
\usepackage{amssymb, amsmath, amsfonts}
\usepackage[english]{babel}
\usepackage{makeidx}
\usepackage{mathrsfs}

\usepackage{graphicx}
\usepackage{comment}
\usepackage{cases}
\usepackage{soul} 
\usepackage[colorlinks=true,breaklinks=true]{hyperref}
\usepackage[dvipsnames, table]{xcolor}
\hypersetup{urlcolor=magenta,
	        citecolor=blue,
	        linkcolor=blue}

\definecolor{Gray}{rgb}{.9,.9,.9}

\newcommand{\bs}[1]{\boldsymbol{#1}}

\begin{document}

\title[]{Probing globular clusters parameters through gravitational wave lensing with stellar-mass black hole binaries}

\author{Sreekanth Harikumar}
\email{sreekanth@camk.edu.pl}
\affiliation{Nicolaus Copernicus Astronomical Center, Polish Academy of Sciences, Bartycka 18, 00-716 Warsaw, Poland.
}

\author{ Abbas Askar}
\affiliation{Nicolaus Copernicus Astronomical Center, Polish Academy of Sciences, Bartycka 18, 00-716 Warsaw, Poland.
}

\author{Micha{\l} Bejger}
\affiliation{Nicolaus Copernicus Astronomical Center, Polish Academy of Sciences, Bartycka 18, 00-716 Warsaw, Poland.
}
\affiliation{INFN Sezione di Ferrara Via Saragat 1 44122 Ferrara Italy}
\author{Marek Biesiada}
\affiliation{National Centre for Nuclear Research, Pasteura 7, 02-093 Warsaw, Poland 
}
\author{Martin Hendry}
\affiliation{SUPA, School of Physics and Astronomy, University of Glasgow, Glasgow G12 8QQ, United Kingdom }

\author{Justin Janquart}
\affiliation{Center for Cosmology, Particle Physics and Phenomenology-CP3, Universit\'e Catholique de Louvain, Louvain-La-Neuve, B-1248, Belgium }
\affiliation{Royal Observatory of Belgium, Avenue Circulaire, 3, 1180 Uccle, Belgium}

\begin{abstract}
Globular clusters (GCs) can act as gravitational lenses for gravitational waves (GWs) in the wave-optics regime, imprinting frequency-dependent signatures on the observed signal. We investigate whether such lensing effects can be used to probe intrinsic properties of GCs, in particular their central velocity dispersion. Modeling GCs as singular isothermal spheres, we simulate lensed GW150914-like signals and perform Bayesian parameter estimation using waveform templates that include both source and lens parameters. We show that the effective lensing mass can be recovered and, when combined with GW sky localization information and GC catalogs, allows for an estimate of the cluster velocity dispersion. For favorable source–lens alignments, the injected values are well recovered within credible intervals. Our results demonstrate that lensed GWs can provide a complementary probe of GC dynamics and motivate searches for such signatures in current and future observations.

\end{abstract}
\maketitle

%
%
{\textit{\textbf{Introduction.}}---}A decade after the first gravitational wave (GW) from a compact binary coalescence (CBC) \cite{LIGOScientific:2016vbw}, the LIGO-Virgo-KAGRA collaboration (LVK) \cite{LIGOScientific:2014pky,VIRGO:2014yos,KAGRA:2013rdx} is shedding light on details of the formation and evolution 
of stellar-mass black holes (BHs). The Advanced LIGO \cite{LIGOScientific:2014pky} and Advanced Virgo \cite{VIRGO:2014yos} detectors have confirmed several hundred GW events to date, with the latest GW transient signals catalogue \cite{LIGOScientific:2025slb, LIGOScientific:2026sit} containing, among others, observations of record high-mass BH binary (GW231123 \cite{LIGOScientific:2025rsn}), loud events (GW250114 \cite{KAGRA:2025oiz} and GW230814 \cite{LIGOScientific:2025cmm}), as well as indications of non-trivial binary formation channels (GW241011 and GW241110 \cite{2025ApJ...993L..21A}). 

With the growing number of GW signals, the first detection of a gravitationally-lensed GW event is not far from reality \cite{Lilan2022,Lilan2019,Mao2018, Smith_2024}. The phenomenon of gravitational lensing is predicted by the theory of general relativity (GR): the presence of a massive object (lens) along the line of sight between the source and the observer distorts the path of the signal. In case of strong lensing, multiple versions of the same signal morphology are produced with a constant shift in phase. These signals are amplified or de-amplified, and arrive at the detector with time delays that depend on the lens mass distribution. 
This phenomenon is routinely observed in the electromagnetic (EM) domain, but the convincing detection of the first gravitationally lensed GW event has not yet occurred    \cite{LIGOScientific:2025cwb,LIGOScientific:2023bwz,LIGOScientific:2021izm,Janquart:2023mvf,Li:2019osa}.

Astrophysical applications resulting from the detection of lensed GWs are manifest. In the strong-lensing limit, where galaxies or galaxy clusters serve as lenses, successful detection of multiple copies of a GW signal allows to study cosmology \cite{Seo:2026eto,Jana:2024uta, Liao:2017ioi,FLRWtest2019}, the lens mass distribution \cite{Seo:2023rjd,Wright_2022}, properties of dark matter \cite{DMviscous2021,DMviscous2022}, and to test theories of gravity \cite{GWspeed2017, Narola:2023viz}. With lenses less massive than galaxies, e.g. globular clusters (GCs), wave optical effects become prominent. Unlike strong lensing, multiple signals do not arise in this regime of lensing, but a frequency dependent amplification of the signal is present \cite{Takahashi:2003ix,Nakamura:1999uwi}. The use of lensed GW signal templates in the wave optics regime allows to establish that the signal is indeed a lensed one and also helps to constrain the nature of the lens itself \cite{Wright_2022,Cheung:2020okf, Mishra:2021xzz}. There is an evidentiary support for the observed high mass event GW231123 \cite{LIGOScientific:2025rsn} to be lensed, due to the non-negligible evidence (see \cite{LIGOScientific:2025rsn,LIGOScientific:2025cwb,Goyal:2025eqo,Shan:2025dcd,Chan:2025kyu} ) found in favour of wave-optical lensing but it is not definitive due to other open questions in data analysis. 

GCs are among the oldest stellar systems in the Universe. They are dense, gravitationally-bound collections of approximately $10^{4}-10^{6}$ stars, with typical half-light radii of a few parsecs, and central densities that can reach up to $\sim 10^{6}\,M_{\odot}\,\mathrm{pc}^{-3}$ \citep{Baumgardt2018}. Their stellar populations are generally metal-poor and are thought to have formed at redshifts $z \sim 2 - 8$, suggesting that they may have played a significant role during the epoch of reionization \cite{Ma:2020yjk}. Consequently, GCs serve as excellent laboratories for studying stellar dynamics, stellar evolution, and the early stages of galaxy formation. Moreover, their spatial distribution and kinematics provide valuable constraints on the mass distribution and structural properties of their host galaxies \cite{Hudson:2014mva}. In addition, GCs serve as important reference points for distance calibration in galaxies \citep{Beasley2024}. They are also prime sites for the dynamical formation and processing of binary black holes (BBHs), enabling these systems to assemble and merge within a Hubble time \citep{spz2000,Benacquista2013,amaro2016,rodriguez2016,askar2017,Mapelli2021}. The GWs emitted by BBHs formed in such clusters can act as probes of their internal properties and dynamical conditions \citep{Hong2018,Romero-Shaw:2020siz,Wu:2025lif}.

Measuring GC masses, typically inferred from their stellar velocity dispersion, $\sigma_v$, is essential for understanding their internal dynamics, formation and subsequent evolution. Current methods of mass determination rely primarily on EM observations, particularly measurements of velocity dispersion and surface brightness profiles \citep{Trager1993,Baumgardt2018}. However, these observations face several limitations, including line-of-sight projection effects \citep{Meylan2002} as well as stellar crowding in the dense central regions of GCs and shot noise from a few luminous giant stars, which can bias velocity-dispersion measurements \citep{bianchini2015,Kimmig2015}. In addition, uncertainties in determining the true cluster center introduce further systematic errors \citep{anderson2010}. These challenges become even more pronounced in core-collapsed clusters or in systems that may host an intermediate-mass BH (IMBH), where the central stellar densities are substantially higher \citep{Noyola2010,Lanzoni2013,Lutzgendorf2012,Lutzgendorf2015,devita2017}.

GCs are found in galaxies of all types. More than 160 have been observed in the Milky Way (MW), over 500 in the Andromeda galaxy \cite{Peacock_2010}, approximately 12,000 in M87, and approximately 70,000 in the Perseus galaxy cluster \cite{Kluge_2025}.
Moreover, the catalog of known GCs will expand significantly with upcoming wide-field surveys, e.g. the LSST \cite{LSSTDESC:2025hol} and Euclid \cite{Euclid:2024yrr}, which will provide precise sky positions, photometry, and distance estimates of GCs \cite{dage2023}. However, measuring their internal kinematics, such as velocity dispersions, remains challenging, and typically requires dedicated spectroscopic follow-up \citep{Puzia2005}.We propose here a new, complementary probe to determine GC properties, in particular their velocity dispersion, by studying a CBC GW event lensed by a GC.  Wave optics effects play a prominent role probing the mass of GCs and hence their velocity dispersion.

In this Letter we first briefly review the basics of lensing theory, and discuss the amplification factor for GCs. The next section contains general characteristics of GCs, in particular those in the vicinity of the MW, and describes the catalogue used in our analysis. Finally, we present our methodology, and conclude with our results.

\begin{figure}[t]
    \centering
    \includegraphics[width=1.15\linewidth]{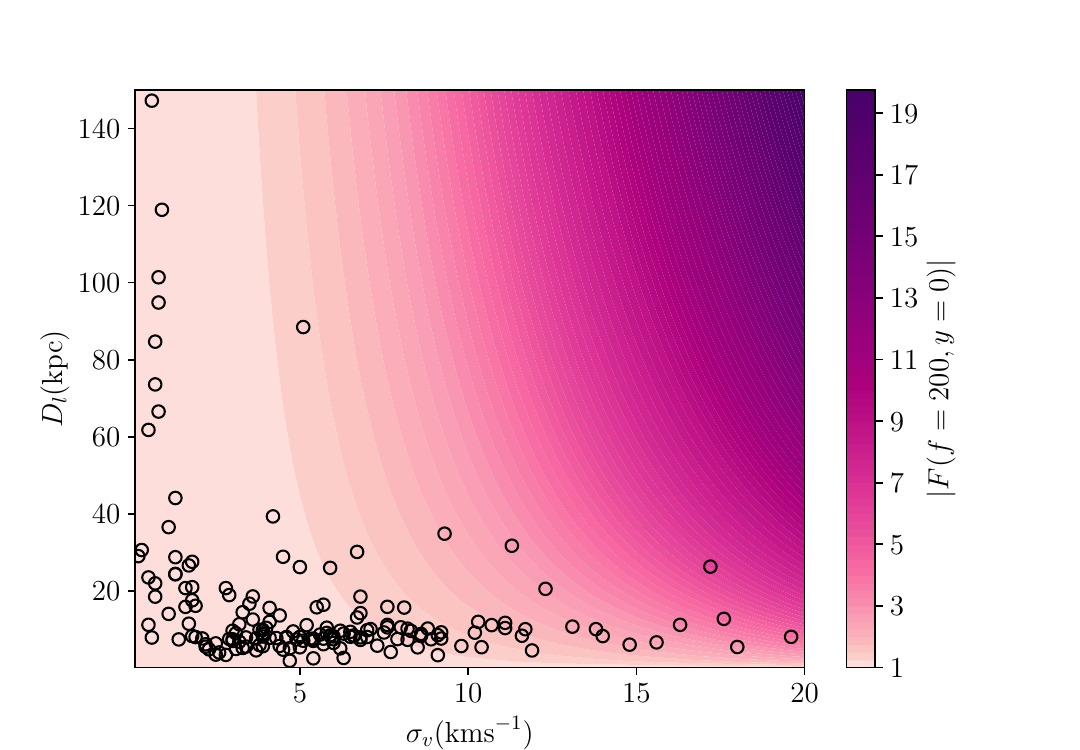}
    \caption{
Color map representing the maximum value of the lensing amplification (Eq.~\ref{sis_amplification}) as a function of the velocity dispersion and the distance to the cluster. Dots display the known GCs catalogued in \cite{Baumgardt2018} and indicate their amplification.
   }
    \label{fig:amp_plot}
\end{figure}

%
%
{\textit{\textbf{Gravitational lensing.}}---}GWs are considered to be propagating in the background spacetime of the lens \cite{Nakamura:1999uwi}. Solving the GW propagation equation in this curved background and assuming a weak gravitational potential leads to the Fresnel-Kirchhoff diffraction integral \cite{Takahashi:2003ix,Nakamura:1999uwi, Gravitational_lenses1992}. This integral, also known as the amplification factor in lensing literature, is defined as the ratio of the lensed and unlensed GW amplitudes in the frequency domain. For an arbitrary lens, the integral takes the following form 
\begin{equation} \label{Fw}
    F(w,\bs y) = \frac{w}{2 \pi i} \int d^2 \bs{x}  e^{  i w T(\bs x,\bs y)},  
\end{equation}
where $w$ is the dimensionless frequency and $\bs {y}$ is the impact parameter in dimensionless form (see also \cite{Takahashi:2003ix,Nakamura:1999uwi,Margherita2023}). The dimensionless frequency can be expressed in terms of the
frequency $f$ of the GW and the effective lens mass $M_{\text{SIS}}$\footnote{\label{footnote_MLz}In most of the GW lensing literature the term redshifted lensing mass, $M_{Lz}$, is used since both lens and source are at cosmological distances. For the case discussed here the lens is at $z_{l} \simeq 0$, hence the notion of the \textit{effective} lens mass.} as  
\begin{equation}\label{eq:dim_freq}
    w = \frac{8\pi G M_{\scriptscriptstyle{\text{SIS}}}f}{c^3}.
\end{equation}
$T(\bs{x}, \bs{y})$ is the Fermat potential that determines the time delay due to lensing; it consists of two contributions: geometric delay and gravitational delay. The Fermat potential in dimensionless form is given by
\begin{equation}
  T(\bs x,\bs y) = \frac{1}{2}|\bs{x} - \bs{y}|^2 - \psi(\bs{x}),  
\end{equation}
where $\psi(\bs x)$ is the effective lensing potential of the lens mass distributions. 
The integral is over the lens plane coordinates $\bs {x}$ and often lacks an analytical solution for the lens models more complex than the point mass or Singular Isothermal Sphere (see below). Therefore, it usually has to be solved numerically \cite{Nakamura:1999uwi, Villarrubia-Rojo:2024xcj, Wright_2022}. 

%
%
{\textit{\textbf{Singular Isothermal Sphere.}}---}The simplest and most popular model of a spherically symmetric self-gravitating system is the Singular Isothermal Sphere (SIS). For simplicity, the GC mass distribution is approximated as a SIS \citep{inagaki1983,Meylan1987}. A distinctive feature of this model is that stars are considered to be isothermal and the velocity dispersion $\sigma_v$ is assumed to be constant. The density profile for the SIS model is given by
\begin{equation}
\rho(r) = \frac{\sigma_v ^2}{2\pi G r^2}.  
\end{equation}
The effective lensing potential for a SIS lens is given by
\begin{equation}\label{sis_lensing_potential}
\psi(x) = |x|.
\end{equation}
The dimensionless frequency takes the form
\begin{equation}
    w = \frac{32\pi^3}{c^5} \sigma_v^4 \frac{D_l D_{ls}}{D_{s}} f\,. 
\end{equation}
Comparing the above equation to Eq.~\eqref{eq:dim_freq}, the effective lens mass $M_{\scriptscriptstyle{\text{SIS}}}$ is identified to be 
\begin{equation}\label{eq:eff_lens_mass}
    M_{\scriptscriptstyle{\text{SIS}}}  = \frac{4\pi^2 \sigma_v^{4} D_l D_{ls}}{c^2 G D_{s}}.
\end{equation}
As the lenses that we consider in this study are situated in the MW, we intentionally omit the redshift term \footref{footnote_MLz} and the mass will be identified as the SIS mass, $M_{\scriptscriptstyle{\text{SIS}}}$. The currently detected CBC sources are much farther ($D_s\simeq 1\,\text{Gpc}$) than the lenses considered in this study ($D_l\simeq 100\,\text{kpc}$). Therefore, the distances in Eq.~\eqref{eq:eff_lens_mass} can be approximated as $D_s \approx D_{ls}$, leading to 
\begin{equation}\label{sis_lensing_mass}
    M_{\scriptscriptstyle\text{SIS}} = \frac{4\pi^2 \sigma_v^{4} D_l} {c^2 G}.
\end{equation} 
 
In the case of the SIS model, the diffraction integral in Eq.~\eqref{Fw} can be calculated analytically, as shown e.g. in \cite{Matsunaga:2006uc}. Hence, the amplification factor can be expressed in terms of dimensionless variables $w$ and $y$ as:  
\begin{eqnarray}\label{sis_amplification}
 F(w) &=& e^{\frac{1}{2}wy^2} \sum_{n=0}^{\infty} \frac{1 + n/2}{n!} (2 w e^{i 3\pi /2 })^{n/2}\ \times \nonumber \\
 &\times& {_1F_1}\Big( -\frac{n}{2};1;\frac{i}{2}w y^2\Big) ,   
\end{eqnarray}
where $_1F_1$ is the confluent hypergeometric function. The SIS is one of the lens models for which Eq.~\eqref{Fw} can be solved exactly and throughout this Letter we restrict our selves to this model. 

\begin{figure}[t]
    \centering
    \includegraphics[width=\columnwidth]{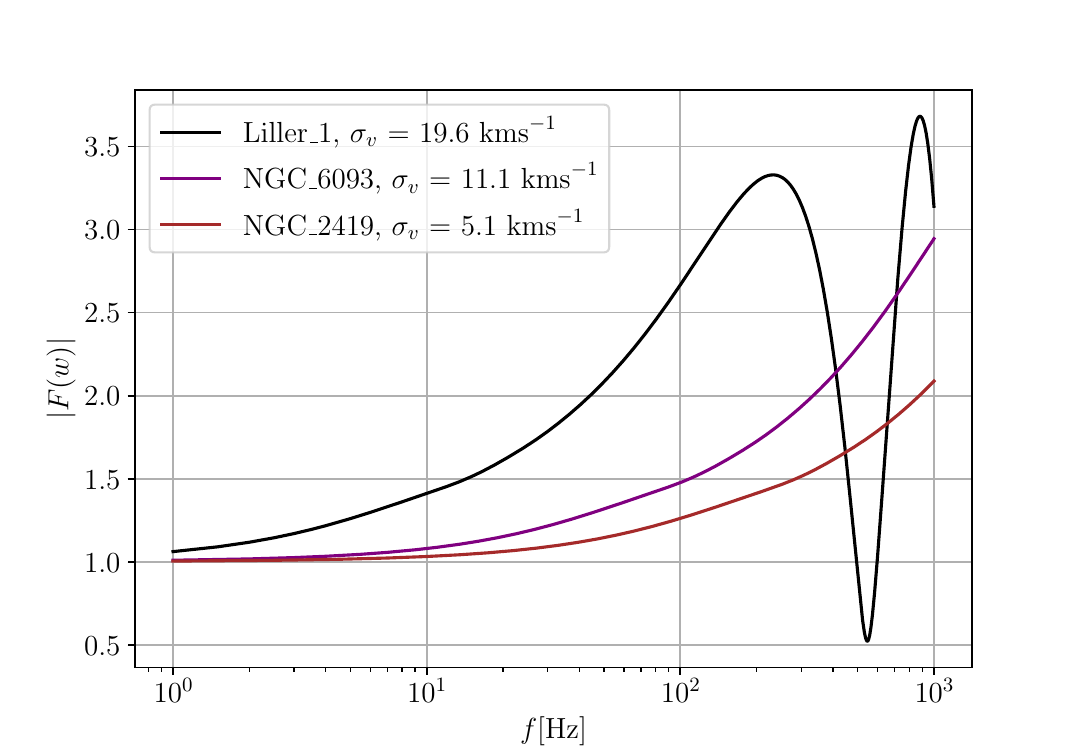}
    \caption{Fresnel - Kirchhoff diffraction integral for GCs with different 1D central velocity dispersion: Liller 1 (black), NGC 6864 (purple) and NGC 6284 (brown), as a function of frequency for a fixed impact parameter $y = 0.3$.}
    \label{fig:amplification_factor}
\end{figure}
%
%
{\textit{\textbf{Catalog of globular clusters.}}---}To demonstrate our methodology, presented in the following section, we limit ourselves to MW GCs whose properties like masses, 1-D central velocity dispersion, distance  to the cluster are well measured in comparison to extra-galactic GCs. Hence, for our analysis we select GCs from the MW catalog. \cite{Baumgardt2018}. GCs in general have a wide range of velocity dispersions, but for the population of 168 known clusters in the MW these lie in the range $\sigma_v \in [ 0.2 ,19.6] \ {\rm km}\;{\rm s}^{-1}$. In general the velocity dispersion varies with respect to the radius of the cluster; however, here we consider only the central velocity dispersion reported in \cite{Baumgardt2018} for the SIS lens model adopted. This value is directly related to the mass of the cluster and the lensing magnification. The maximum amplification -- as characterized by the condition $ F(w, y=0)$ -- that a MW GC can achieve acting as a lens, at fixed frequency ($f_{gw} = 200 \ {\rm Hz}$), is shown in Fig.~\ref{fig:amp_plot}.
It can be seen from Fig.~\ref{fig:amp_plot} that, as expected, GCs with larger velocity dispersions display larger amplificationfor a fixed frequency and impact parameter. A significant fraction of the clusters in the catalogue \cite{Baumgardt2018} have $\sigma_v < 7.5 \ {\rm km}\; {\rm s}^{-1}$, for which $F(f,y)$ remains close to unity across the frequency band of interest (here illustrated at $f_{gw}=200$). Since SNR is evaluated by integrating $|F|^2$ over the signal in the detector band, such clusters would not produce an appreciable enhancement in SNR.

%
%
{\textit{\textbf{Methodology.}}---}
To assess whether the velocity dispersion of a GC can be recovered from a lensed GW signal, we perform an injection study. We simulate GW150914-like signals, lens them with a SIS amplification factor corresponding to the chosen GC, and inject them to a Gaussian noise \cite{KAGRA:2013rdx} generated from the O4 power spectral densities \footnote{For LIGO detectors we used the PSD found in \url{https://dcc.ligo.org/public/0165/T2000012/002/aligo_O4high.txt} and Virgo we used PSD found in \url{https://dcc.ligo.org/public/0165/T2000012/002/avirgo_O4high_NEW.txt}} (PSDs) of a three-detector network comprising LIGO Hanford (H), LIGO Livingston (L), and Virgo (V). The signals are subsequently recovered using lensed waveform templates that incorporate the parameters of both the CBC source and lens. From the recovered posterior distributions, we obtain sky-localization constraints and compute the corresponding credible area on the sky. This region is then cross-matched with the MW GC catalog to identify potential lensing GCs. Once a candidate GC is identified, its known distance from the catalog is then used in Eq.~\eqref{eq:eff_lens_mass} to invert the recovered $M_{\scriptscriptstyle{\text{SIS}}}$ posterior and estimate the corresponding velocity dispersion of the cluster.

 We assume the simulated  events be observed by a three detector network consisting of Hanford (H), Livingston (L) and Virgo (V). The injection parameters were obtained from the median values of posterior samples for the IMRPhenomXPHM \cite{Pratten:2020ceb} approximant in the GWTC-2.1 catalog \cite{LIGOScientific:2021usb}. 
 Given the unlensed waveform, the corresponding lensed waveform is obtained by multiplying Eq.~\eqref{sis_amplification} with it. We choose three different GCs from the MW catalog \cite{Baumgardt2018} with distinct values for the central velocity dispersion $\sigma_v$; these GCs were chosen according to the strength with which they could amplify a GW signal -- noting that the amplification is proportional to the lensing mass, as given by Eq.~\eqref{sis_lensing_mass}, through the dimensionless frequency $w$ which depends of the velocity dispersion $\sigma_v$. The selected GCs are: \textit{Liller-1} -- a cluster with a larger velocity dispersion  $\sigma_v = 19.06 \ \mathrm{km\, s}^{-1}$; \textit{NGC6093} -- a cluster with an intermediate velocity dispersion $\sigma_v = 11.10 \ \mathrm{km\, s}^{-1}$ and \textit{NGC 2419} -- a cluster with a lower velocity dispersion $\sigma_v = 5.10 \ \mathrm{km\, s}^{-1}$. To study the recoverability of the velocity dispersion for these GCs
 we perform 20 different injections, varying only the dimensionless impact parameter in the range $[0.001,5]$ for each cluster. As the amplification is a function of $y$ the strength of the amplification, and hence the signal to noise ratio of the lensed signal, will decrease as the impact parameter increases. The GCs are well separated in the sky and hence for each cluster the true GW source position is placed behind the chosen cluster, where the exact position of the cluster is obtained from \cite{Baumgardt2018}. 
 
 It is to be noted that location of the source varies with the choice of impact parameter $y$ as $\alpha' = \alpha$ and $\delta' = \delta +{y \xi_0}/{D_l}$, where $\alpha$ and $\delta$ are the right ascension and declination of the GC respectively, coinciding with the $\alpha'$ and $\delta '$ of the GW source for a perfect source-lens alignment ($y=0$). The Einstein radius of the SIS lens is denoted by $\xi_0$. These injected signals are recovered with a SIS lensed waveform implemented in the GWLENS \cite{GWLENS} python package and the source and lens parameters are inferred using the python-based Bayesian Inference tool Bilby \cite{bilby_paper,bilby_pipe_paper} and the nested sampling technique implemented in dynesty \cite{dynesty2020}. 
\begin{figure}[t]
    \centering
\includegraphics[width=1.0\linewidth]{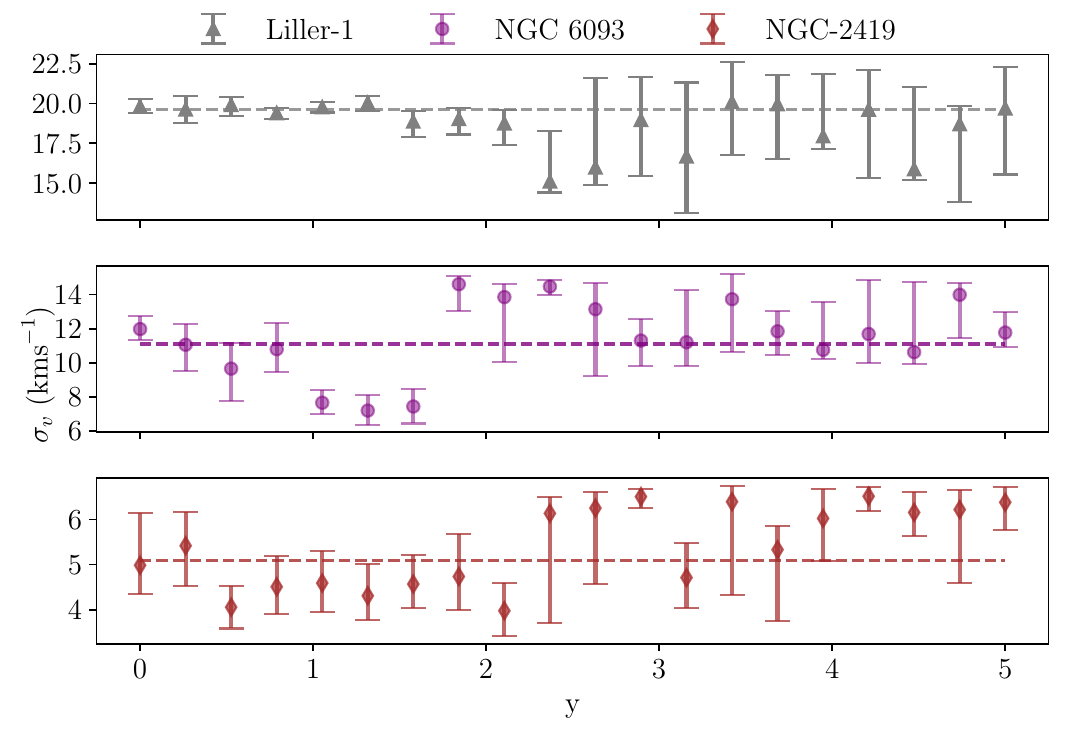}
    \caption{Median values of the velocity dispersion $\sigma_v$ recovered from the lensed signals for different as a function of impact parameter $y$, with error bars indicating the 68\% credible interval. Different colours show different different clusters: Liller 1 (gray triangles), NGC-6093 (purple circles), NGC-2419 (brown diamonds).  The horizontal dotted lines represent their injected values. }
    \label{fig:sigma_for_3clusters}
\end{figure}
\noindent For each of the injections for the chosen three clusters we recover the 15 binary parameters of the IMRPhenomXPHM approximant and the lens parameters including the SIS lensing mass $M_{\scriptscriptstyle\text{SIS}}$ and the dimensionless impact parameter $y$. From the recovered sky localization posteriors, a 90\% credible region for the sky-localization area is constructed  and cross-matched with all known GCs in the MW catalog \cite{Baumgardt2018} to identify the cluster responsible for lensing. When more than one cluster falls within the credible region, we rank candidates by the posterior probability enclosed at the cluster position from the maximum-posterior sky location and associate the event with the top ranked cluster. For the identified cluster the velocity dispersion is estimated from the following expression 
\begin{equation}\label{eq:vd_sampling}
    \sigma_v = \Big[ \frac{c^2 G }{4 \pi^2} \frac{M_{\scriptscriptstyle{\text{SIS}}}}{D_l}\Big]^{1/4} ,
\end{equation}
where the distance to the cluster $D_l$ is obtained from the EM catalog and the lensing mass is recovered from the lensed signal parameter estimation.

\begin{figure}[htbp]
  \centering

  \begin{minipage}{0.48\textwidth}
    \centering
    \includegraphics[width=\linewidth]{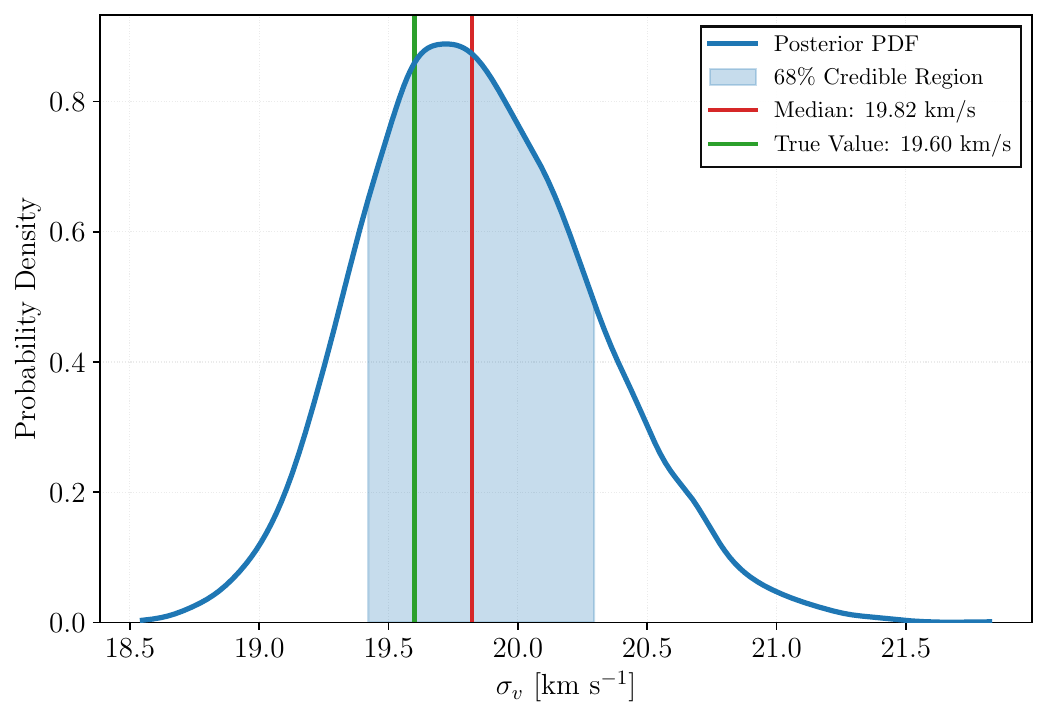}
  \end{minipage}%
  \hfill

  \begin{minipage}{0.48\textwidth}
    \centering
    \includegraphics[width=\linewidth]{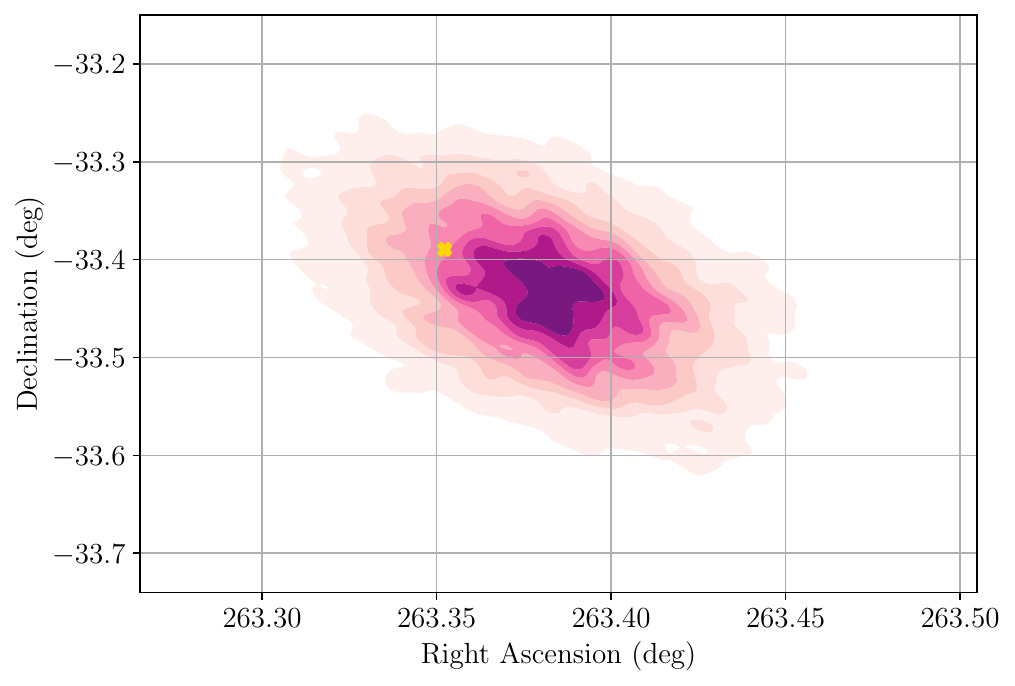}
  \end{minipage}

  \caption{Velocity dispersion (top panel) and sky localization (bottom panel) and for Liller 1 and $y=0.001$.}
  \label{fig:skyloc_sigma}
\end{figure}

%
%
{\textit{\textbf{Results.}}---}We find that for the chosen set of 20 different impact parameters investigated for each cluster, the injected cluster lie within the recovered 90\% credible region for the sky-localization area. Moreover, from cross-matching of the recovered sky area with the entire MW catalog \cite{Baumgardt2018}, we found that the injected cluster was identified correctly . The angular distance $D_l$ of the identified GC is picked from the catalog to recover the velocity dispersion using Eq.~\eqref{eq:vd_sampling}. For the cluster with higher velocity dispersion, \textit{Liller-1}, the recovered probability distribution peaks at the injected value $\sigma_v = 19.6 \ \mathrm{km\, s}^{-1}$, with median value $\sigma_v = 19.7 \ \mathrm{km\, s}^{-1}$,  as shown in Fig.~\ref{fig:skyloc_sigma}. The true value lies within the $68 \%$ credible interval, indicating that the inference successfully recovers the velocity dispersion for impact parameter $y<2$. For subsequent injections with higher impact parameters, $\sigma_v$ is also recovered; however the uncertainty of the inference is seen to increase and the recovered median values of the velocity dispersion are found to be shifted away from the true value due to random noise realisations (see Fig.~\ref{fig:sigma_for_3clusters}). We also notice the anti-correlation between the recovered parameters $M_{\scriptscriptstyle{\text{SIS}}}$ and $y$ for the impact parameter $y \sim 1$ (in NGC-6093) as reported for a point mass lens in \cite{Mishra:2023ddt}, which has been verified with zero noise injection to discard variations due to random statistical fluctuations. For the chosen three different GCs (Liller-1, NGC 6093 and NGC-2419) the recovered velocity dispersions at different impact parameters and their uncertainties are shown in Fig.~\ref{fig:sigma_for_3clusters}. This can be understood as follows: as $y$ increases the amplification decreases, weakening the imprint of the lens on the signal. This leads to poor sampling of the lensing mass $M_{\scriptscriptstyle\text{SIS}}$ for larger impact parameters. The same trend can be seen for other two clusters as shown in Fig.~\ref{fig:sigma_for_3clusters}.

%
%
{\textit{\textbf{Conclusions.}}---}This Letter demonstrates that a GW source lensed by a GC can be used to investigate and constrain intrinsic properties of the GC, namely its velocity dispersion. Inferring the velocity dispersion allows to measure the mass of the GC, which is essential to understand their internal dynamics, formation and subsequent evolution. Additionally, it could also indicate the possible presence of IMBHs at its core. The presented methodology can also be used to confidently identify a GW event as a (micro)lensed GW signal, if one of the MW GCs happens to be the lens. The probability for a MW GC to act as a lens is estimated as $\tau \sim 10^{-8}$ (see the supplemental material and \cite{Ubach:2025dob}), highlighting the potential of next-generation detectors to discover lensed GW events by MW GCs and probe globular cluster physics; it will therefore be important to understand how lensing by GCs will affect the recovery of source parameters, including biases \cite{Mishra:2023ddt} and degeneracies with e.g. higher modes \cite{Ezquiaga:2022nak}, precession\cite{Liu:2023emk}, eccentricity \cite{Mishra:2025dpa}, beyond-GR effects \cite{Wright:2024mco,Mishra:2023vzo} etc. Furthermore, there is a high probability that an extragalactic GC will act as the lens, for which there may not be GC catalog information available. Therefore, one would have to carry out a more extensive parameter estimation including both the source and lens distance as well as the lens velocity dispersion. As the lensing mass $M_{\rm{\text{SIS}}}$ depends on the both $\sigma_v$ and the angular diameter distance to the lens $D_l$, a degeneracy between them is expected.
Improved extragalactic GC catalogs could help break this degeneracy, highlighting the multi-messenger potential of the project. \\
Where the lens does belong to a GC catalog, accurate identification of the correct GC in the catalog depends on the accuracy of the sky-localization of the signal. A global network of at least three GW detectors with comparable sensitivity would, therefore, be an important requirement to enhance the method's further applicability and efficacy.
To demonstrate our methodology, we used here a SIS lens model; consequently, the information that we could recover is limited to the velocity dispersion of the cluster. However, the parameters of more realistic models, such as cored isothermal spheres, Plummer spheres, lowered-isothermal King models \citep{king1966}, and other tidally truncated models, could also be recovered using our methodology. A detailed study with these lens models in the MW and extra-galactic GCs will be investigated in a follow up work. \\
%
%
\\
{\textit{\textbf{Acknowledgements.}}--}
SH is grateful to Anirudh Nemmani, Anuj Mishra, Mirek Giersz and Sudhagar Suyamprakasam for their useful comments and discussions. 
We gratefully acknowledge Polish high-performance computing infrastructure PLGrid (ACK Cyfronet AGH) grant no. PLG/2025/018667, and CAMK PAN computing resources. This work was partially supported by the Polish National Science Centre (NCN) grants no. 2021/43/B/ST9/01714, 2023/50/A/ST9/00579, 2024/55/D/ST9/02585, and 2025/56/C/ST9/00480, and by the UK Science and Technology Facilities Council (Grant Ref. ST/L000946/1).  J.J. acknowledges support from the Fonds de la Recherche Scientifique (FNRS) and the Royal Observatory of Belgium. For the purpose of Open Access, the author has applied a CC BY public copyright licence to any Author Accepted Manuscript (AAM) version arising from this submission.

\bibliography{refs}

\section*{Supplemental Material\label{sec:supp}}

\subsection*{Optical depth}
Using the formula from \cite{Paczynski1994}, the optical depth for the GCs can be written as 
\begin{equation}
\tau = \frac{\Sigma(r)}{\Sigma_{\text{cr}}},.
\end{equation}
Assuming an SIS model, the critical surface mass density for a GC can be written as $\Sigma(r) = {\sigma_v ^2}/({2 G r})$ and the critical surface mass density is given by
$\Sigma_{cr} = ({c^2}/{4 \pi G})({D_s}/({D_l D_s}))$, where $r$ is the distance from the cluster center. This distance $r$ is related to $D_l$ by $r = \phi D_l$, where $\phi$ is the angular distance from the cluster center. Using the above relations, we obtain
\begin{figure}[b]
\centering
\includegraphics[width=1.0\linewidth]{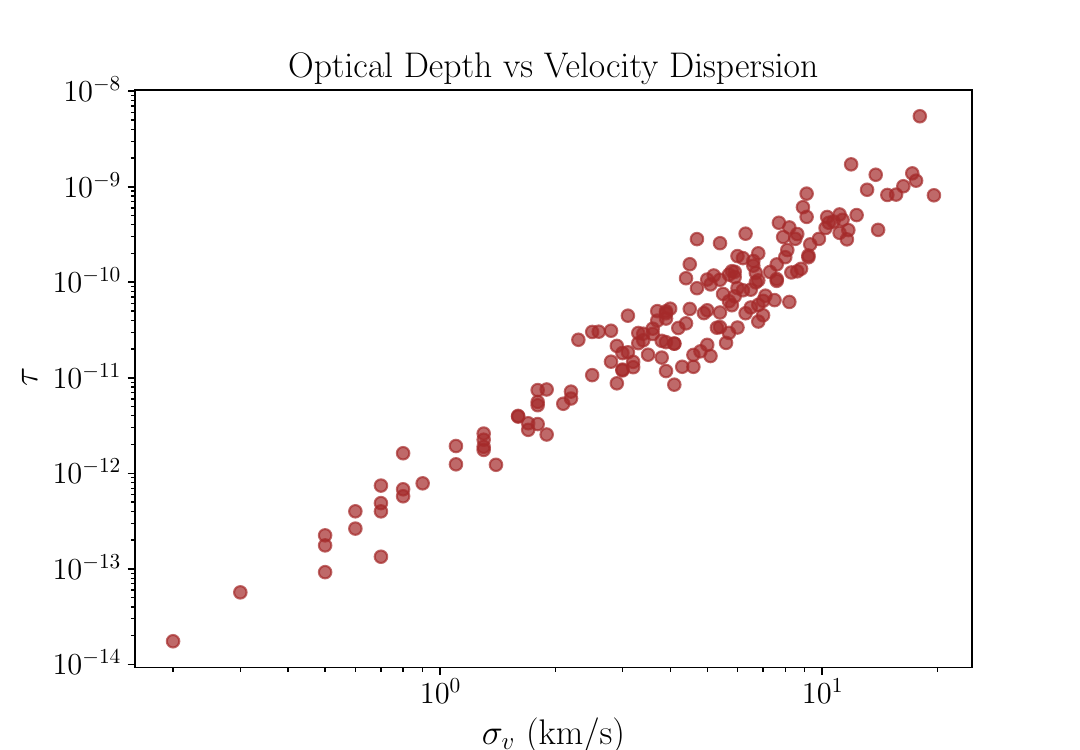}
\caption{Optical depth for GCs in the MW as a function of their measured central velocity dispersion $\sigma_v$}
\label{fig:optical_depth2}
\end{figure}
\begin{equation}
\tau = \frac{\sigma_v^2}{c^2} \frac{2 \pi}{\phi}\left( 1 - \frac{D_l}{D_s} \right)\,.
\end{equation}
For a lens in MW and the source at cosmological distances $D_s \gg D_l$. Hence one can approximate the above expression as 
\begin{equation} \label{tau_phi}
\tau(\phi) \approx \frac{\sigma_v^2}{c^2} \frac{2 \pi}{\phi}\,. 
\end{equation}
The above equation describes the optical depth for lensing at the radial-angular distance $\phi$ from the center of the GC on the sky. In order to calculate total optical depth for the cluster, one has to integrate it up to the (angular) radius of the cluster $\phi_{GC}$, which can be chosen as e.g. the core radius. Hence, the total optical depths is \begin{equation} \label{tau_phi2}
\tau = \int_0^{\phi_{GC}} \tau(\phi) 2 \pi \phi d\phi \approx 4 \pi^2\frac{\sigma_v^2}{c^2} \phi_{GC}, 
\end{equation}
which yields $\tau = 1.23 \, 10^{-11}\,\left(\frac{\sigma_v}{10\,km/s} \right)^2 \left(\frac{\phi_{GC}}{1'} \right)$. Using the MW globular cluster catalog \cite{Baumgardt2018}, one can calculate the lensing probability for the MW globular clusters as 
\begin{equation}
    \tau_{\scriptscriptstyle{tot}} = \sum_{i}^{N} \tau_i\,.
\end{equation}
The optical depth has been found to be $3.16 \times 10^{-8}$. It is to be noted that for the calculation of angular radius we used the tidal radius of each GC computed from the catalog.
\end{document}